\pdfoutput=1
\documentclass[a4paper,fleqn]{cas-dc}

\usepackage[numbers]{natbib}

\usepackage{booktabs}
\usepackage{tikz}
\usepackage{pgfplots}

\usepackage{CJK}
\usepackage{algorithm}
\usepackage{algorithmicx}
\usepackage{algpseudocode}
\usepackage{amsfonts,amssymb}
\usepackage{listings}
\usepackage{color}
\usepackage{graphicx}
\definecolor{dkgreen}{rgb}{0,0.6,0}
\definecolor{gray}{rgb}{0.5,0.5,0.5}
\definecolor{mauve}{rgb}{0.58,0,0.82}
\usepackage{diagbox}
\usepackage{CJK}
\def\tsc#1{\csdef{#1}{\textsc{\lowercase{#1}}\xspace}}
\tsc{WGM}
\tsc{QE}
\tsc{EP}
\tsc{PMS}
\tsc{BEC}
\tsc{DE}

\begin{document}
\let\WriteBookmarks\relax
\def\floatpagepagefraction{1}
\def\textpagefraction{.001}
\shorttitle{Modeling Programs Hierarchically with Stack-Augmented LSTM}
\shortauthors{Fang et~al.}

\title [mode = title]{Modeling Programs Hierarchically with Stack-Augmented LSTM}                     

\author[1]{Fang Liu}
\ead{liufang816@pku.edu.cn}

\author[1]{Lu Zhang}
\ead{zhanglucs@pku.edu.cn}

\author[1]{Zhi Jin}
\ead{zhijin@pku.edu.cn}
\cormark[1]

\address[1]{Peking University, Beijing 100871, China }

\cortext[cor1]{Corresponding author}

\begin{abstract} Programming language modeling has attracted extensive attention in recent years, and it plays an essential role in program processing fields. Statistical language models, which are initially designed for natural languages, have been generally used for modeling programming languages. However, different from natural languages, programming languages contain explicit and hierarchical structure that is hard to learn by traditional statistical language models. To address this challenge, we propose a novel Stack-Augmented LSTM neural network for programming language modeling. Adding a stack memory component into the LSTM network enables our model to capture the hierarchical information of programs through the PUSH and POP operations, which further allows our model capturing the long-term dependency in the programs. We evaluate the proposed model on three program analysis tasks, i.e., code completion, program classification, and code summarization. Evaluation results show that our proposed model outperforms baseline models in all the three tasks, indicating that by capturing the structural information of programs with a stack, our proposed model can represent programs more precisely.
\end{abstract}

\begin{highlights}
\item Stack is suitable for handling the hierarchical structures
\item Adding a stack into LSTM enables our model to capture the hierarchical information
\item Our model can capture the long-term dependency in the programs
\item Our model can represent programs precisely, outperforms baselines in three tasks
\end{highlights}

\begin{keywords}
Programming language modeling  \sep Hierarchical structure \sep Deep learning \sep Software engineering
\end{keywords}

\maketitle

\section{Introduction}
Programming language modeling is a hot research topic in both software engineering and artificial intelligence. Learning and analyzing programming languages provide a way of understanding programs, which is helpful in many program analysis tasks, such as code completion \cite{hellendoorn2017deep}, code summarization \cite{iyer2016summarizing}, code clone detection \cite{wei2017supervised}, etc. Similar to natural languages, programming languages also contain predictable statistical properties \cite{hindle2012naturalness}. For example, \textit{for i in range (...)} is used frequently in many Python source code files. Thus, programming languages can be modeled by statistical language models. As we know, statistical language models are widely used in natural language processing. These models have shown great performance on many NLP tasks, such as machine translation \cite{brown1990statistical}, speech recognition \cite{jelinek1975design}, etc. In recent years, statistical language modeling approaches, including deep learning based models, have been applied to modeling programming languages \cite{hindle2012naturalness,tu2014localness,raychev2014code,raychev2015predicting,raychev2016learning,bielik2016phog,liu2016neural}. Hindle et al. \cite{hindle2012naturalness} discovered the predictable properties of programming languages and proposed a code completion model based on n-gram model. Raychev et al. \cite{raychev2015predicting} built a statistical model for programs to predict program properties based on conditional random fields (CRFs). Recently, deep learning based statistical models are applied to modeling programming languages. Liu et al. \cite{liu2016neural} proposed a code completion model based on LSTM network. Li et al. \cite{Li2018Code} proposed a pointer mixture network to address the out-of-vocabulary (OoV) issue in code completion.

Despite some similarities between programming languages and natural languages, there are also some differences \cite{pane2001studying}, of which the most important one is that programming languages contain explicit and hierarchical structure. There are many code blocks of different granularity in programs, such as statements, loops, methods, classes, etc. These code blocks are always nested and different code blocks compose to form meaningful and functional programs. In some large programs, the nesting can be really deep and makes the programs hard to understand. The hierarchical structure is closely related to the syntax and semantics of programs, which is quite important in program modeling. From a practical point of view, integrating a hierarchical structure into a neural-network based language model for program language modeling may be important for the following reasons: 
\begin{itemize}
    \item to produce an accurate and hierarchical representation for the program, which will be helpful for many downstream program analysis tasks, especially for those tasks that focus more on the program's structural information. 
    \item to model the compositional effects of the program, and thus can help with long-term dependency problem \cite{Bengio09} by providing shortcuts for gradient back-propagation.
    \item to improve generalization of the model through biasing the neural networks towards performing tree-like composition operations and thus can achieve good performance on many downstream tasks.
\end{itemize}

However, in most of the current source code modeling approaches  \cite{hindle2012naturalness,tu2014localness,liu2016neural,bhoopchand2016learning,Li2018Code}, programs are represented as flatten sequences, for example, token sequences or AST node sequences. Then they built statistical language models, including N-gram based models or neural network based models, to model the source code sequences. During the model's learning process, the element in the program sequence is processed one by one without considering the program's hierarchical structure. The model cannot realize where a new code block begins or ends during this process. Besides, several AST or graph based models have been proposed to represent programs \cite{allamanis2017learning,alon2018general} with the purpose of capturing the structural information in programs. They use ASTs or enrich ASTs by introducing the data-flow edges to build graphs as the representation of programs, and build neural networks to process the AST or graph. These models focus more on the data-flow information and are used for variable usage or naming tasks. They also needed to adding data-flow edges into the AST or extract explicit AST paths, which is tedious. Besides, in the above models, they can only produce the representations for complete code snippets, which cannot be used for the scenario of the incomplete inputs, such as code completion task. In this paper, we aim to build a general statistical language model for programs based just on the programs (AST), which can be used for many downstream program analysis tasks, including code completion, program classification, code summarization, etc.

In this paper, we propose a statistical language model to learn programming languages based on a Stack-Augmented LSTM neural network. Associating the recurrent neural network with a memory unit can improve the network by capturing more information in data sequences \cite{graves2014neural,joulin2015inferring,kumar2016ask}. We adopt Long-Short-Term-Memory (LSTM) \cite{hochreiter1997long} neural network as the base model and strengthen the network by adding a stack as a memory component. The characteristic of the stack is well suitable for handling the hierarchical structure of programs. The PUSH and POP operations can be used to record the hierarchical structure of programs. When the model starts to process a new code block, we PUSH the hidden state that stores the program's contextual information into the stack. After processing this entire code block, we POP off the stack and update the hidden state of the network to get the general information of the code block. When the code blocks are nested, the internal (lower level) code blocks will be handled first. Then this information can be used to understand the external (higher level) code blocks. Adding a stack enables the network to discover the hierarchical dependencies of source code, which further allows the model capturing longer dependency in the input program. Our work provides a new perspective on how to incorporate hierarchical information into program representations.

We experimentally test the proposed model on three program analysis tasks, including code completion, classifying program by their functionality, and code summarization. The results show that our model outperforms baseline models on all these tasks, which demonstrate the effectiveness of our model. 
Our main contributions are as follows:
\begin{itemize}
\item We propose a novel neural-network based language model aiming at modeling the hierarchical structure of the programs. We strengthen the LSTM network with a stack to store and restore the contextual information depending on the program's structure, which enables the model to capture the structural information of programming languages, and further allows the model capturing the long-term dependency in programs.
\item We apply the proposed language model to three program analysis tasks. Evaluation results demonstrate that our model outperforms all the baseline models, which indicates that by capturing the hierarchical structure of programs, our model could understand programs better and represent programs more precisely.
\end{itemize} 

\noindent\textbf{Paper Organization} ~ The remainder of this paper is organized as follows. We introduce our motivation in Section \ref{motivation}. Section \ref{related_work} highlights some work related to this paper. Then we provide background knowledge on statistical language models in Section \ref{language_model} for a better understanding of our proposed model. In Section \ref{model}, we introduce our proposed model. Section \ref{exp} shows the evaluation details. In Section \ref{res}, we present the experimental results and make careful analysis of the results. Then we discuss some limitations and threats to validity of our model in Section \ref{discussion}. Finally, we conclude our paper in Section \ref{conclusion}.

\section{Motivation}\label{motivation}
\begin{figure}
\setlength{\abovecaptionskip}{0cm}
\centering\includegraphics[width=8.5cm]{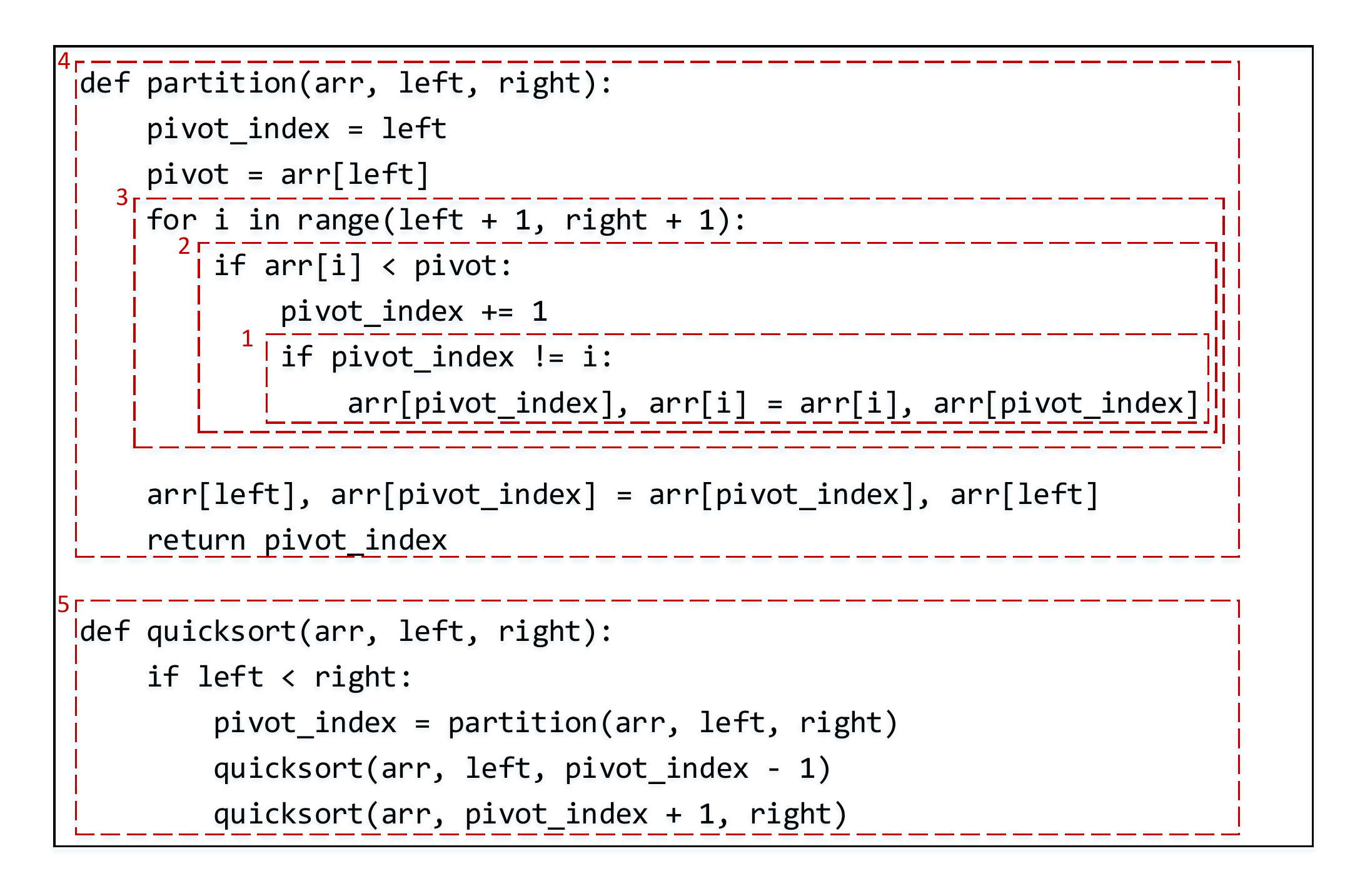} 
\caption{Quick-sort algorithm of Python} 
\vspace{-0.3cm}
\label{Fig:code_block}
\end{figure}

Programs cannot be simply viewed as a sequence of tokens because of their explicit and hierarchical structure. As we have mentioned before, code blocks are nested in programs. Take the \emph{quick-sort} algorithm in Figure \ref{Fig:code_block} as an example, \emph{FOR} loops, \emph{IF} statements, and function definitions are all code blocks. These code blocks are nested. The whole \emph{quick-sort} algorithm is also a code block, and it can also be nested in a larger code block. Programs are hierarchically organized like this. The nested code blocks lead to the hierarchical structure of programs, which makes programs hard to learn by traditional statistical language models. 

Understanding the hierarchical structure of languages is essential for statistical language models, especially in programming languages. Remembering all the tokens in the above algorithm is hard for humans. Instead, they can recognize the hierarchy of the programs and compose the small code blocks to form meaningful code snippets in their brains. First, they will figure out the functionality of the lower-level code blocks and then exploit this information to understand the higher-level code blocks. When they read the algorithm in Figure \ref{Fig:code_block}, they will first understand the \emph{FOR} loops (code block 3) and its sub-code blocks (code block 1 and 2) in the \emph{partition} function. Then they can figure out the functionality of the \emph{partition} function (code block 4), that is, choosing a pivot and rearranging the array to make sure that everything to the left of the pivot is less than the pivot and everything to the right is greater than the pivot. Then they will figure out the functionality of the \emph{quick-sort} function (code block 5), which recursively calls on partition function until there is nothing left to partition. When reading the \emph{quick-sort} function (code block 5), they do not care the implementation details of the \emph{partition} function. Instead, they only remember the functionality of it. That is the way humans understand programs in most cases. An efficient and suitable statistical language model for modeling programming languages should have the ability to think like a human. In this paper, we propose a Stack-Augmented LSTM network based language model to discover the hierarchical structure of source code as well as capture longer-term dependency during the programming language modeling process. 

\section{Related Work}\label{related_work}
There has been much interest in programming language modeling. Different models have been proposed, and these models mainly fall into two categories, i.e., sequential models and structural models. In recent years, memories have been generally used to strengthen the ability of neural networks. Inspired by these models, we adopt a stack as a memory component to augment our model. In this section, we introduce some related work on these models.

\subsection{Sequential Models for Programming Languages}
Sequential models view code as a sequence of elements, usually tokens or AST nodes. These models predict sequences by sequentially generating each element. N-gram model is a widely used sequential model, which is effective for capturing local statistical dependencies in input sequences. Hindle et al. \cite{hindle2012naturalness} discovered the predictable properties of programming languages and proposed a code completion model based on the n-gram model. Then, many n-gram based models are proposed for modeling programming languages \cite{nguyen2013statistical,tu2014localness,hellendoorn2017deep}. Hellendoorn et al. \cite{hellendoorn2017deep} have introduced an improved n-gram model that addresses the challenges of modeling source code by considering the special features of code, including unlimited vocabulary, nested scope, locality, and dynamism. In recent years, as the amount of open source code increases, deep recurrent neural network based language models have been applied to learning source code and have made great progress. The hidden state in the recurrent neural network can encode longer range dependencies than n-gram models. More importantly, RNN models are also powerful to learn distributed representations of words, which can represent more information in the programs. Liu et al. \cite{liu2016neural} proposed a Vanilla LSTM model for code completion. They represent programs as AST node sequences and built an LSTM model to predict each node based on the previous nodes. Bhoopchand et al. \cite{bhoopchand2016learning} proposed an RNN model with a sparse pointer mechanism for modeling the program token sequences, aiming at capturing long-range dependencies in the programs. Then, Li et al. \cite{Li2018Code} proposed a pointer mixture network to address the OoV issue in the code completion task. They represent programs as the ASTs and flatten each AST in in-order depth-first
traversal to produce the node sequence as the input of the language model. Their model has achieved state-of-the-art results in the code completion task, which is used as a baseline in this paper. In these approaches, programs are represented as token sequences or AST node sequence, and they built sequential models to model these sequences. These models did not pay attention to the hierarchical structure of the program.

\subsection{Structural Models for Programming Languages}
Structural models view code at the level of ASTs or graphs. In contrast to sequential models, structural models focus on learning the structural information of programs. Mou et al. \cite{mou2016convolutional} proposed a tree-based convolutional neural network (TBCNN) for modeling programs' structural information. In their model, a set of tree-based convolution kernels are adopted to sliding over the entire AST for extracting structural information. However, it can only be used for producing the representation of a complete program, which cannot be used in tasks like code completion, where the inputs are incomplete program sequences. Alon et al. \cite{alon2018general} represented a program using paths between nodes in the AST to enable their learning model to leverage the structured nature of source code. Using AST paths as the representation for programs helps the model capture the data-flows and learn how the variables are used. They evaluated their approach on the tasks of predicting variable names, method names, and types. Recently, Alon et al. \cite{alon2019code2vec} extend this approach for method name predicting task. Allamanis et al. \cite{allamanis2017learning} enriched the AST by introducing the data-flow edges, and they build a graph neural network to process them. Their model focused more on the data-flow information, and they evaluated their model on “variable naming” and “variable misuse” tasks, where the data-flow information plays an important role. In the above models, they concerned more about the data-flow information, and these models are only used for variable (method) usage or naming tasks. These models can only produce the representations for complete code snippets, which cannot be used for the scenario of the incomplete inputs. Besides, they needed to extract explicit AST paths or adding data-flow edges into the AST, which are tedious. The motivation and scenario of their model are different from ours. We hope to build a general language model for programs based just on the programs, which can be used for general program analysis tasks, including code completion, program classification, code summarization, etc. Our model learns the hierarchical structural information from the original AST and focuses on the global statistical information of programs instead of local information like variables, which is helpful for understanding the programs from hierarchical and global perspectives. 
 
\subsection{Memory-augmented Models}
Memory components have been introduced to strengthen recurrent networks for capturing non-sequential features in recent years. Graves et al. \cite{graves2014neural} developed a neural turing machine that adds an external memory component to a neural network, which was capable of learning simple algorithms such as copying, sorting, and associative recall from input and output examples. Joulin et al. \cite{joulin2015inferring} associated a recurrent network with a stack and a list as memories to enable the network to solve pattern recognition problems, which involved some form of counting and memorization. They used gating mechanisms as learnable controllers to decide the operations between PUSH, POP, and NO-OP based on a probability distribution. Inspired by the above work, we build a statistical language model based on a Stack-Augmented LSTM network for learning programming languages. Different from the above models, when to PUSH or POP is deterministic based on the AST structure of programs in our model. In this way, we enable the model to make full use of the tree structural information of programs. We utilize the PUSH and POP operations of the stack to trace the beginning and end of code blocks, which helps the model to capture the hierarchical features in the programming languages and represent programs more precisely. 

\section{Statistical Language Models for Programming Languages}\label{language_model}
In this section, we present some background knowledge on statistical language models and how to applied these models in programming language modeling. Programming languages are kind of languages that contain predictable statistical properties \cite{hindle2012naturalness}. Many code snippets occur frequently in programming languages. Therefore, programs can be learned by statistical language models which can capture the statistical patterns in programs by assigning occurrence probabilities to a sequence of tokens. Given previously seen context tokens $t_{1},t_{2},...t_{t-1}$, the model can predict the probability of the next token $t_{t}$, i.e.,:
\begin{equation}\label{eq1}
p(t_{t}|context)=p(t_{t}|t_{1},t_{2},...,t_{t-1})
\end{equation}

\subsection{N-gram based Models}
The probabilities in equation \eqref{eq1} are hard to estimate because the number of the context sequences $t_{1},t_{2},...,t_{t-1}$ is tremendous. N-gram model based on the Markov assumption is proposed aiming at addressing this challenge. In the n-gram model, the probability of a token is dependent only on n-1 most recent tokens, i.e.,:
\begin{equation}
p(t_{t}|t_{1},t_{2},...,t_{t-1})=p(t_{t}|t_{t-n+1},...,t_{t-1})
\end{equation}
N-gram model is the most commonly used statistical language model in natural language processing, and many n-gram based models have been applied to program analysis tasks like code completion \cite{hindle2012naturalness,tu2014localness}. These models have been proved effective for capturing the repetitive regularities in the source code. However, programming languages are verbose, where data dependency can be very long. For example, a variable defined at the beginning of a program can be used at the end of the program. Hence, n-1 tokens of context may not contain enough information for modeling programming languages.

\subsection{Deep Learning based Models}
\begin{figure}
\centering\includegraphics[width=6cm]{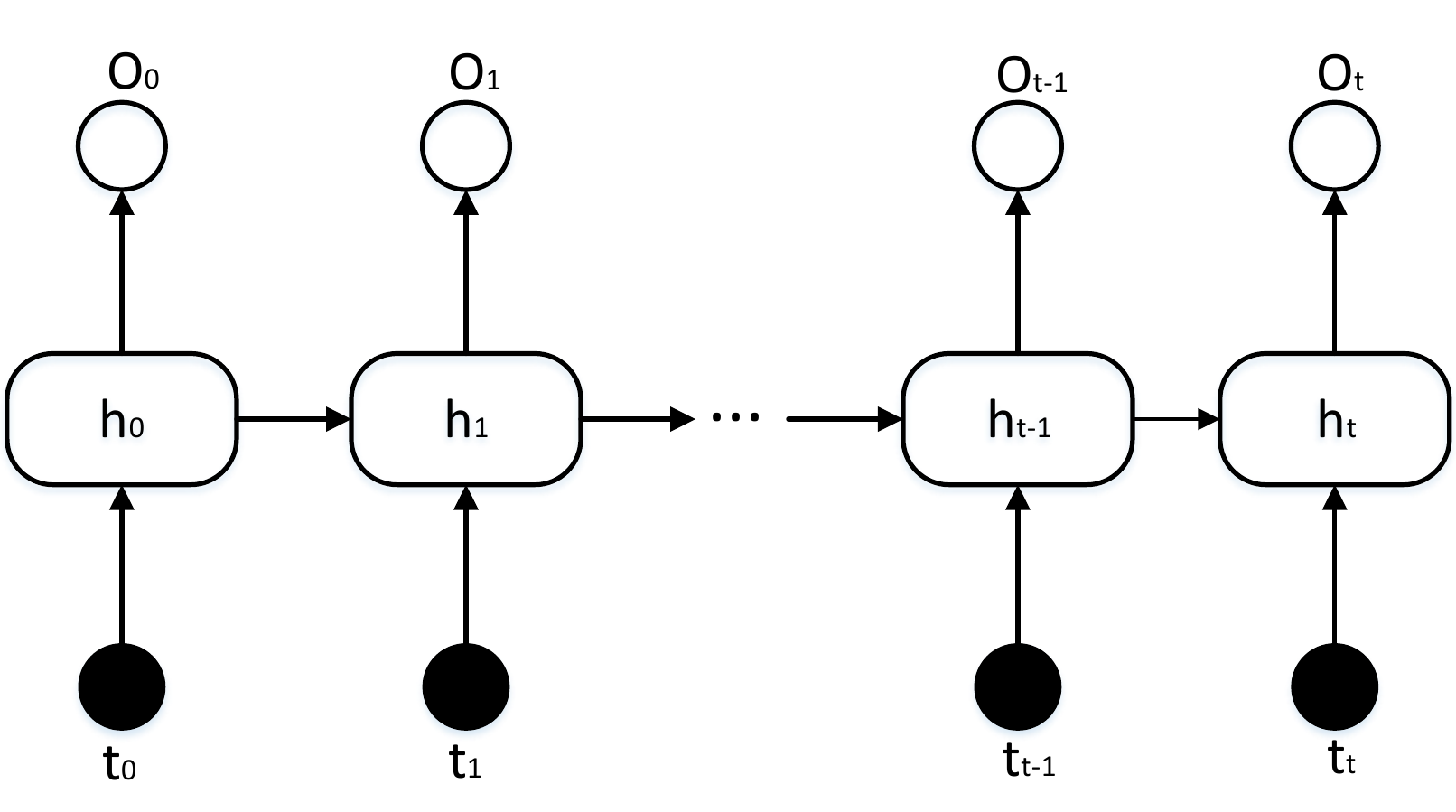}
\caption{\label{Fig:rnn} Architecture of RNN.}
\vspace{-0.3cm}
\end{figure}

In recent years, deep learning techniques have made great progress in many fields. Deep recurrent neural network based statistical language models have also shown great performance in modeling both natural languages and programming languages. Mikolov et al. \cite{mikolov2010recurrent} proposed a recurrent neural network based language model. By using recurrent connections, information can cycle inside these networks for a long time. Figure \ref{Fig:rnn} shows an RNN network architecture for modeling programs. Tokens in a program are fed into the network successively. The hidden state $h_{t}$ in the network represents the information of the inputs in previous time steps $t_{1},...,t_{t-1}$. The hidden state is fed into the output layer to get the predicted token. RNN models can loosen the fixed context size and capture longer dependencies than the n-gram model.

However, the vanishing gradient problem prevents standard RNN from learning long-term dependencies. LSTM is an extension for RNN, which was proposed by \cite{hochreiter1997long} as a solution to the vanishing gradient problem. With the powerful gate mechanism, LSTM can remember and forget information from context selectively. Thus, it is widely used for modeling sequential data, and it has been used for learning programming languages \cite{bhoopchand2016learning,liu2016neural,Li2018Code}.

\section{The Proposed Model}\label{model}
In this section, we present our Stack-Augmented LSTM (SA-LSTM) model in detail. We build our model based on the LSTM network. As aforementioned, programs contain hierarchical structure, which is hard to capture by the vanilla LSTM network. We augment the LSTM network by adding a stack to capture the structural information in programs.

\begin{figure*}
\setlength{\abovecaptionskip}{0cm}
\centering\includegraphics[width=14cm]{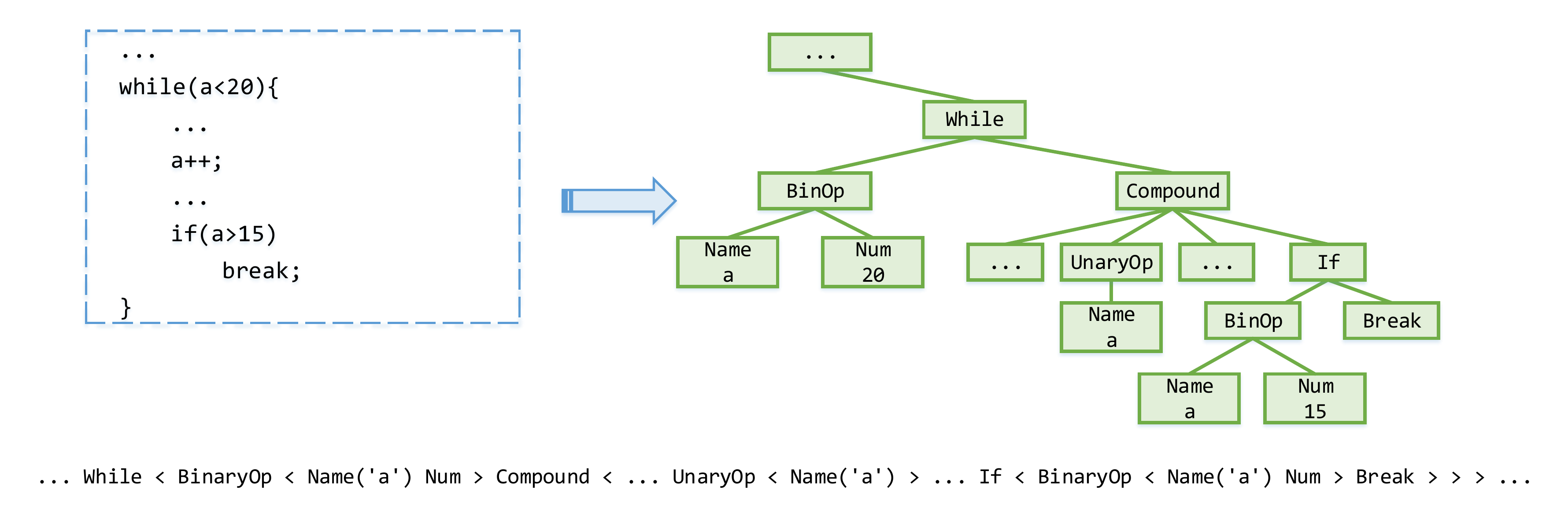} 
\caption{While loop in C, its corresponding AST, and flattened AST.} 
\label{Fig:while}
\vspace{-0.3cm}
\end{figure*} 

\subsection{Program Representation} 
In recent program language modeling research, programs are mainly represented in three ways, i.e., token-based representation, AST-based representation, and Graph-based representation. In token-based program representation \cite{bhoopchand2016learning,hindle2012naturalness}, programs are tokenized into token sequences, and each token is represented as a real-valued vector. Compared to AST and Graph based representation, representing the programs as token sequences preserves the natural order of typing, and even if the programs are incomplete or contain bugs, they can still be tokenized. In AST-based program representation \cite{liu2016neural,raychev2016probabilistic,Li2018Code,alon2019code2vec}, programs are represented as ASTs. In programming languages, each program can be parsed into a unique AST since programming languages have an unambiguous context-free grammar. ASTs are widely used for processing programs to extract the syntax and structure of programs. In AST-based representation, programs are first be parsed into ASTs, and then ASTs are traversed to node sequences or paths between AST leaf nodes are extracted. Each node of the AST denotes a construct occurring in the source code, where the non-leaf nodes correspond to non-terminals, and leaf nodes correspond to terminals. In programs, non-terminals are the type of code blocks. For example, \emph{Assign, If, For, While}, etc. These non-terminals constitute the skeleton of programs, which contain the structural information. Terminals in programs can be a variable name, a string, a number, an operator, etc. In Graph-based program representation \cite{allamanis2017learning,cvitkovic2018open}, they enrich ASTs by introducing the data-flow edges to build graphs as the representation of programs, and build neural network to model the graphs. These models focus more on the data-flow information and are used for variable usage or naming tasks.

In our model, we aim at modeling the structural information of programming languages. Thus, we need to find a proper way to preserve the hierarchical structure of the programs in the program's representation. Then we can build a model to learn this information from programs. The hierarchical structure of the source code can be stored in the tree structure of the AST. Thus, we use AST to represent programs. Take the \emph{While} loop (in C) and its corresponding AST in Figure \ref{Fig:while} as an example, the structure of the code is stored in the tree's hierarchical structure integrally. Each node of the tree denotes a construct occurring in the code snippet. For example, the conditional statement is stored in the \emph{BinOp} node and its sub-trees, and the body of the \emph{While} loop is stored in the \emph{Compound} node and its sub-trees. After parsing source code into AST, we need to serialize it to produce the node sequence as the input of the language model. To preserve the hierarchical structure of the tree, we adopt a new way to flatten the AST. We traverse the tree in a depth-first order. During the traversal, we use $\langle$ and $\rangle$ to surround the sub-tree of each non-leaf node, which corresponds to code blocks. Therefore, each code block begins with $\langle$ and ends with $\rangle$. When serializing the AST in Figure \ref{Fig:while} using this method, we can get the sequence shown at the bottom of the figure. In this way, the hierarchical structure is preserved, and ASTs can be flattened without losing the structural information. 

\subsection{Stack-Augmented LSTM Programming Language Model}
\subsubsection{Model Architecture}
LSTM has been widely used for modeling sequential data. The hidden state of the LSTM network can store the information of the input data. Hence, we adopt an LSTM as our base model. However, LSTM lacks the ability to capture the structural information of data. As we have mentioned before, programs contain explicit and hierarchical structure. To address the limitation of LSTM in modeling programs, we build a Stack-Augmented LSTM language model (SA-LSTM) for programming languages, which can learn the hierarchical information of programs.

\begin{figure*}
\centering\includegraphics[width=13cm]{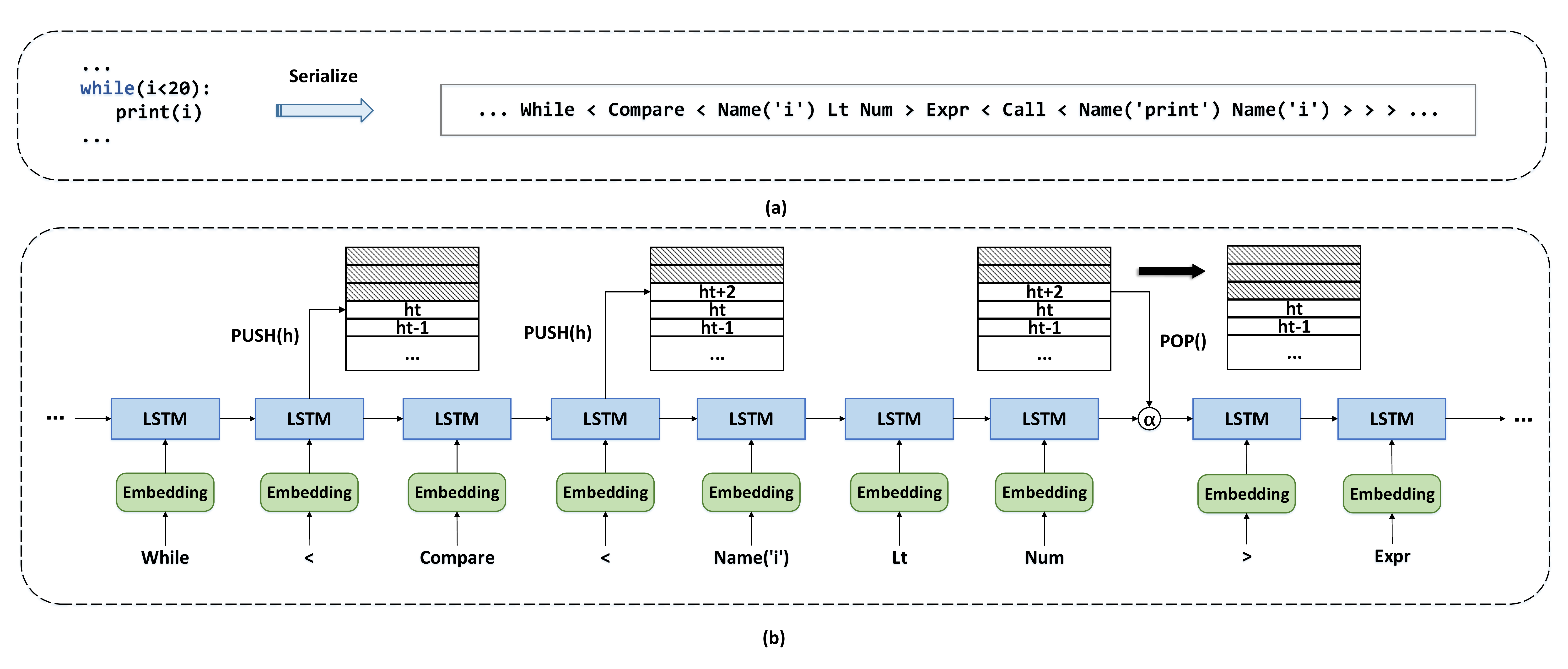} 
\caption{SA-LSTM model: (a) The sample code snippet of the input program and its sequential representation. (b)The overall structure of SA-LSTM model.}
\label{Fig:model}
\end{figure*} 

Figure \ref{Fig:model} shows the structure of SA-LSTM. Figure \ref{Fig:model}a is an example of \emph{While} loop in Python and its sequential representation. We take this sequence as the input of the SA-LSTM model in this example. Figure \ref{Fig:model}b is the overall structure of SA-LSTM. The tokens (AST nodes) in the sequence are fed into the model one by one. During the model's training process, the stack's PUSH and POP operations are performed on the hidden state to enable SA-LSTM to store and restore contextual information. The PUSH and POP operation is deterministic based on the AST structure of programs. When a new code block is fed into the model, we PUSH the hidden state $h_{begin}$ into the stack. Then the hidden state is updated. For example, when the $\langle$ next to While is fed into the model, which means that the model begins to learn the code block of \emph{While} loop. After the model has read all the tokens in the code block and reached the end of it, we POP off the stack to get the hidden state $h_{begin}$ which is stored at the beginning of the code block. Then $h_{begin}$ and the hidden state of the previous step $h_{t-1}$ are combined to obtain the code block's information with the function $\alpha$. The value of hidden state $h_{t}$ is updated as equation \ref{eq:salstm}, where $x_{t} \in \mathbb{D}$ represents the input vector at time step t. $V_{start}$ and $V_{end}$ represent the vector of start and end symbol of a code block, respectively. $h_{t}$ represents the hidden state of the LSTM at time step t. $c_{t}$, $i_{t}$, $o_{t}$ and $f_{t}$ represent the cell and three gates, i.e., input gate $i_{t}$, output gate $o_{t}$, and forget gate $f_{t}$. At every time step, our model first decides how to handle the previous hidden state $h_{t-1}$ and gets the context hidden state $h_{context}$. Then we use $h_{context}$ and the input vector to compute the current hidden state $h_{t}$, which is the same as the standard LSTM's hidden state updating process. We use $h_{t}=lstm(x_{t},h_{context})$ as a simplified representation of this process.

\begin{equation}
\begin{split}
\label{eq:salstm}
&h_{context}=\left\{
\begin{array}{lr}
h_{t-1}, \ push(h_{t-1})\ \ \ \ \ \ \ x_t=V_{start} \\
\alpha (pop(), h_{t-1}) \ \ \ \ \ \ \ \ \ \ \ \   x_t=V_{end}\\
h_{t-1} \ \ \ \ \ \ \ \ \ \ \ \ \ \ \ \ \ \ \ \ \ \ \ \ Other\ cases
\end{array}
\right. \\
&f_{t}=\sigma_{g}(W_{f}x_{t}+U_{f}h_{context}+b_{f}) \\
&i_{t}=\sigma_{g}(W_{i}x_{t}+U_{i}h_{context}+b_{i}) \\
&o_{t}=\sigma_{g}(W_{o}x_{t}+U_{o}h_{context}+b_{o}) \\
&c_{t}=f_{t}\cdot c_{t-1}+i_{t} \cdot \tanh (W_{c}x_{t}+U_{c}h_{context}+b_{c})\\
&h_{t}=o_{t} \cdot \tanh c_{t}
\end{split}
\end{equation}

The stack memory has two primary operations:
\begin{itemize}
\item PUSH: Push the hidden state $h$ onto the stack.
\item POP: Return and remove the top element of the stack.
\end{itemize}

Since the hidden state preserves the information of input data, we perform the PUSH and POP operations on hidden state according to the hierarchical structure of the programs, which enables SA-LSTM to store and restore contextual information depending on the programs' structure. In this way, the model is endowed with the ability to discover structural dependencies of programs. The detailed algorithm will be introduced in the next section.

\subsubsection{Algorithm}
The hierarchical structure information of source code is stored in AST. After processing the AST using our method, each code block begins with “$\langle$” and ends with “$\rangle$”. Take the following serialized AST sequence as an example:\\
\textbf{Module $\langle$ While \underline{$\langle$} Compare $\langle$ Name('i') Lt Num $\rangle$ Expr $\langle$ Call $\langle$ Name('print') Name('i') $\rangle$ $\rangle$ \underline{$\rangle$} $\rangle$ }\par
In the above sequence, \textit{While} corresponds to a non-leaf node in AST. Tokens in \textit{While} loop are surrounded by the underlined angel brace. When the model reads \underline{$\langle$}, it means that the model is going to learn the tokens in the \emph{While} code block. When \underline{$\rangle$} is read, it means that the model has read all the tokens in the \emph{While} loop.
The algorithm for SA-LSTM is presented in Algorithm \ref{Algorithm:1}. $V_{start}$ and $V_{end}$ represent the vector of start and end symbol of a code block, i.e., $\langle$ and $\rangle$. There are three cases:\\
\renewcommand{\algorithmicrequire}{\textbf{INPUT:}}
\begin{algorithm}[t]
\caption{SA-LSTM}
\label{Algorithm:1}
\begin{algorithmic}[1]
\Require $x$\
\State $Initialize\ model\ parameters$
\For{t=1 to numsteps}
\If {$x_t=V_{start}$}  
\State $PUSH(h_{t-1})$;
\State $h_t \gets lstm(x_t,h_{t-1})$
\ElsIf {$x_t=V_{end}$} 
\State $h_{begin} \gets POP()$ 
\State $h_{context} \gets \alpha(h_{begin},h_{t-1})$
\State $h_t \gets lstm(x_t,h_{context})$
\Else
\State $h_t \gets lstm(x_t,h_{t-1})$
\EndIf		      
\EndFor 
\end{algorithmic}
\end{algorithm}
(1) Lines 3-5 in Algorithm \ref{Algorithm:1}: Input token is $\langle$, i.e., the model accesses a non-leaf node and is going to learn a new code block, we push the hidden state into the stack. Then the hidden state is updated with the \emph{lstm} function given in the previous subsection. \\
(2) Lines 6-9 in Algorithm \ref{Algorithm:1}: Input token is $\rangle$, i.e., the model has read all the tokens in the code block and reaches the end of it, we pop off the stack to get the hidden state $h_{begin}$ which is stored at the beginning of the code block. Then $h_{begin}$ and $h_{t-1}$ are combined to extract the key information of the code block with function $\alpha$. $h_{begin}$ is the hidden state produced at the beginning of the code block, and $h_{t-1}$ is the hidden state of the previous time step, i.e., the hidden state produced at the end of the code block. We store this valuable information into $h_{context}$. Then we use $x_{t}$ and $h_{context}$ as the input of \emph{lstm} function to update the hidden state. \\
(3) Lines 10-11 in Algorithm \ref{Algorithm:1}: Other cases, we just adopt the same process as the standard LSTM network to update the hidden state.

$\alpha$ is an arbitrary trainable function, which is designed for extracting the important contextual information from $h_{begin}$ and $h_{t-1}$ when a complete code block has been accessed by the network. We can implement $\alpha$ in different ways. We have tried the following options based on the purpose of extracting valuable contextual information from $h_{begin}$ and $h_{t-1}$. 
\begin{itemize}
\item Fully-connect: Concatenate $h_{begin}$ and $h_{t-1}$ into a new vector, then fed it to a fully connected layer:\\
$h_{context}=fc(concat(h_{begin},h_{t-1}))$
\item Max-pooling: Perform max-pooling operation between the $h_{begin}$ and $h_{t-1}$: \\
$h_{context}=maxpooling(h_{begin},h_{t-1})$
\item Summarization: Take the hidden state of the previous time step $h_{t-1}$ as a “summary” of the context, and use it as a new input vector for the network. Then set current state to $h_{begin}$ and execute \emph{lstm} function get $h_{context}$: \\
$h_{context}=lstm(h_{t-1},h_{begin})$
\end{itemize}

\section{Evaluation}\label{exp}
\subsection{Research Questions}
To evaluate our proposed model, we address the following research questions:
\begin{itemize}
    \item \textbf{RQ1: }How does our proposed model perform when compared with baseline models? 
    \item \textbf{RQ2: }How do different $\alpha$ functions impact the performance of our proposed model? 
    \item \textbf{RQ3: }Is our proposed model effective for learning the hierarchical structure of programs?
    \item \textbf{RQ4: }Is our model effective for capturing long-term dependency in programs?
\end{itemize}

\subsection{Tasks}
We evaluate our proposed model on three program analysis tasks, including code completion, classifying programs by functionalities, and code summarization. These three tasks occupy important places in software engineering and program analysis. The performance of these tasks can be improved with a better understanding of the programs' hierarchical structure.

\subsubsection{Code Completion}
Code completion is an essential feature of integrated development environments (IDEs) that is extensively used by programmers. It can speed up the process of software development by suggesting the next probable token based on existing code. Code completion can be considered as a prediction task, and every token in the program is predicted according to their context. Statistical language models are widely adopted on this task. By discovering the hierarchical feature of programs, our model might make a better decision in predicting the next token.

\subsubsection{Program Classification}
Classifying programs by their functionalities occupies an important place in software engineering \cite{mou2016convolutional}. For instance, by detecting the certain patterns of the source code, programmers could discover bugs of the source code, or detect the code clone. This task can also be considered as a prediction task. Given a program, predict which category it belongs to. With a better statistical language model that could learn the hierarchical structure of programs, preciser representations can be produced, which might be helpful for the classification task.

\subsubsection{Code Summarization}
Natural language summaries of source code are valuable in many software applications \cite{iyer2016summarizing}. The natural language descriptions of the source code can help programmers understand the programs, which can be used for code searching, code learning tutorials, etc. It is necessary to understand the programs' structure in this task. Hence, our proposed statistical language model can be used for this task to produce a structured representation for programs. Based on this representation, high-quality summarizations might be produced.

\subsection{Metrics}
To train and evaluate the models, we need to define metrics to measure the model's performance. We use different metrics for different tasks as shown in Table \ref{Tab:metric}:
\begin{table}
  \begin{center}
  \setlength{\tabcolsep}{7mm}{
    \caption{\label{Tab:metric}Metrics for different tasks.}
    \begin{tabular}{lc}
      \hline
      \textbf{Tasks} & \textbf{Metrics} \\
      \hline
       Code Completion & Accuracy, MRR \\
       Program Classification & Accuracy \\
       Code Summarization & BLEU-4, METEOR \\
      \hline
    \end{tabular}
    }
  \end{center}
\end{table}

\subsubsection{Accuracy} For code completion and program classification tasks, we adopt accuracy as the metric. These two tasks are prediction tasks. In the code completion task, given the previous tokens, the model will produce a probability distribution of the tokens in the pre-defined vocabulary and choose the token with the highest probability. While in the program classification task, the model produces a probability distribution of the categories and chooses the category with the highest probability. Hence, we use accuracy to measure the performance of the model on these two tasks. Accuracy evaluates the proportion of the correctly predicted samples, and the calculation formula is shown in equation \eqref{eq4}. \\ 
\begin{equation}\label{eq4}
Accuracy = \frac{\# \ Correct\ predicted\ samples}{\# \  Total\ samples}
\end{equation}

\subsubsection{MRR} Mean Reciprocal Ranking (MRR) is a statistic measure for evaluating any process that produces a list of possible responses to a sample of queries, ordered by the probability of correctness. The reciprocal rank of a query response is the multiplicative inverse of the rank of the first correct answer: 1 for first place, 1/2 for second place, 1/3 for third place, and so on. In this paper, we only compute MRR among the top-10 suggestions, where the correct suggestion appears out of the top-10 suggestions is considered as wrong predictions and its MRR is 0. For this metric, larger values indicate the better performance of the model. We adopt MRR as another metric for code completion task, and this matric has been used in previous code completion models \cite{hellendoorn2017deep,karampatsis2019maybe}.

\subsubsection{BLEU}
BLEU (bilingual evaluation understudy) \cite{papineni2002bleu} is an automatic algorithm for evaluating the quality of natural language text, and it measures how close a candidate sequence is to a reference sequence. It is a widely used measure for machine translation. Generally, BLEU measures the n-gram overlap between the texts produced by models and the reference texts. It is computed as:

\begin{equation}\label{eq5}
BLEU = BP \cdot \exp(\sum_{n-1}^N w_n\log p_n)
\end{equation}

where $p_n$ is the precision of n-grams, that is, the ratio of n-grams in the candidate that are also in the reference:
\begin{equation}\label{eq6}
p_n = \frac{\# \text{n-grams}_{overlap}}{\# \text{n-grams}_{candidate}} \text{ for n=1,...,N}
\end{equation}

where N is the maximum number of grams we consider, and we set N to 4. Each $w_n=\frac{1}{N}$ is the weight of $p_n$. $BP$ is a brevity penalty which penalties the short candidates:
\begin{equation}\label{eq7}
BP = \left\{
\begin{array}{lr}
1  \ \  &  {\text{if} \ c > r} \\
e^{(1-r/c)} \ \  &  {\text{if} \ c \le r} \\
\end{array}
\right. \\
\end{equation}
where $r$ is the length of the reference sequence, and $c$ is the length of the candidate sequence. We use BLEU to evaluate the generated summaries for code summarization task. 

\subsubsection{METEOR} METEOR (Metric for Evaluation of Translation with Explicit ORdering) \cite{banerjee2005meteor} is another metric for the evaluation of machine translation output. It is based on the harmonic mean of unigram precision and recall, where recall weighted higher than precision. Different from BLEU, METEOR considers several features that are ignored in other metrics, including stemming and synonymy matching, along with the standard exact word matching. METEOR can produce a good correlation with human judgment. We also adopt METEOR as a metric to evaluate the generated summaries for code summarization task. 

\subsection{Datasets, Baselines, and Training Details}
\subsubsection{Code Completion}

\noindent\textbf{Datasets} ~ For code completion task, we conduct our experiments on two real-world datasets covering two different programming languages: C and Python. The programs in C datasets come from the OJ system \cite{mou2016convolutional}. The programs in Python datasets are collected from GitHub \footnote{https://github.com/LiuFang816/SALSTM\_py\_data}. We select the projects that have at least 5 stars in 2007. After cleaning the data, there are 51745 C source code files and 37986 Python source code files. We randomly split the data for training, validation, and testing at a ratio of 8:1:1. 

\noindent\textbf{Baselines} ~ In code completion, we predict every token in the program based on the previous tokens. In this task, the input program are incomplete, where models like TBCNN \cite{mou2016convolutional} and other AST path based models \cite{alon2018general,alon2019code2vec} can not be used. Thus we compare SA-LSTM with the following two language modeling approaches: Vanilla LSTM \cite{liu2016neural} and Pointer Mixture Network \cite{Li2018Code}. In Pointer Mixture Network, pointer mixture network is adopted to enable the model to generate next token from either the global vocabulary or the local context, which can ease the OoV problems. Besides, they also propose a new parent attention for the AST-based code completion, which makes use of the structural information in AST. Their model achieves state-of-the-art results in code completion task. 

\noindent\textbf{Training Details} ~ We adopt two layers LSTM network with the hidden unit size of 200 in SA-LSTM and the other two baseline models. The vocabulary sizes for C and Python datasets are set to 1000 and 5000, respectively. We set the unrolling length of the LSTM to 400, which corresponds to the length of the input programs (AST nodes). For the programs which are shorter than 400, we add paddings to the end of those programs. For Pointer Mixture Network, the size of the attention window is 50, which is set the same as in their paper. The embeddings for each node in AST are randomly initialized, and the embedding size is set to 200. We use Adam optimizer with the base learning rate of 1e-3 for our model and the two baseline models. We train our model for 7 epochs, Vanilla LSTM for 8 epochs and Pointer Mixture Network for 8 epochs. 

\subsubsection{Program Classification}

\noindent\textbf{Datasets} ~ For the program classification task, we use C code in the OJ system \cite{mou2016convolutional} as the dataset. The number of categories is 104. In each category, there are 500 programs, and programs in the same category have the same functionality. 

\noindent\textbf{Baselines} ~ In program classification task, the programs are classified by their functionalities, where the representation for the program is of great importance. In this paper, we aim to build a model for capturing the hierarchical structural information of the programs and produce a precise representation for the programs as well. We compare our model with vanilla LSTM and two structure-based model: Tree-LSTM \cite{tai2015improved} and TBCNN \cite{mou2016convolutional}. Tree-LSTM is a tree-based LSTM network that aims at capturing the structural information of the tree. Nodes in the tree are processed bottom-up. In TBCNN, a set of tree-based convolution kernels are adopted to sliding over the entire AST for extracting structural information. The representations for the AST nodes are learned by a coding criterion \cite{peng2015building}.

\noindent\textbf{Training Details} ~ For TBCNN \cite{mou2016convolutional}, since we use the same dataset as them, we directly report the results in their paper. For vanilla LSTM, Tree-LSTM and SA-LSTM, we set the hidden unit size to 600, where the parameter size is comparable with TBCNN. The hidden state of the last time step in networks contains the information of a complete program, so that we use it as the representation and feature of the program for classification. In these models, the embeddings for each node in AST are randomly initialized. The vocabulary size is 1,000, and the embedding size is 600. The length of the programs (AST nodes) are set to 600, which can achieve the best performance according to the experimental finding in section \ref{long_dependency}. We use Adam optimizer with a base learning rate of 1e-3. We train our model for 8 epochs, Vanilla LSTM, and Tree-LSTM for 10 epochs. For TBCNN, we directly copy the results from their paper. 

\subsubsection{Code Summarization}

\noindent\textbf{Datasets} ~ In this task, We perform experiments on Java dataset. Java dataset is from Hu et al. \cite{Hu18Summarizing}, which contains 69,708 pairs of Java methods and summaries. The Java projects are collected from GitHub, and the methods and corresponding Javadoc comments (serve as the summaries) are extracted. The programs in the datasets are split into training, valid and testing sets in proportion with 8 : 1 : 1 after shuffling the pairs. The average lengths of Java methods and comments are 99.94 and 8.86, respectively.

\noindent\textbf{Baselines} ~  In code summarization task, we apply our model as the encoder to encode the programs and produce a structured representation for programs. We compare our model with the following approaches.
\begin{itemize}
    \item seq2seq+attention: A seq2seq model with attention mechanism \cite{bahdanau2014neural}, where the encoder and decoder are both LSTM networks. 
    \item TL-CodeSum \cite{Hu18Summarizing}: A code summarization model that uses API knowledge learned in a different but related task. The Java dataset used in this task is from this paper.
    \item Code2vec \cite{alon2019code2vec}: A structural code representation model that uses AST path to represent programs. 
    \item Hybrid2Seq+Attn+DRL (HAD) \cite{wan2018improving}: A code summarization model that also considers the structural information of source code. They use both token sequence and AST to represent source code, and present a hybrid embedding for combining these two representations. Besides, they also adopt a deep reinforcement learning framework, named actor-critic network, to cope with the exposure bias issue existing.
\end{itemize}

In our model, we adopt the encoder-decoder framework, and use SA-LSTM as the encoder to produce the representations of programs. Besides, we also apply the attention mechanism \cite{bahdanau2014neural} in this task. Attention mechanism was designed for allowing a model to automatically pay attention to parts of a source sentence that are relevant to the predicted word. It was generally applied to sequence-to-sequence tasks such as machine translation \cite{bahdanau2014neural}, question-answering \cite{li2015diversity}, text summarization \cite{chopra2016abstractive}, as well as code summarization \cite{iyer2016summarizing}. With the attention mechanism, the decoder could ``attend" to different parts of the source sequence at each time step by giving different weights to input tokens. Thus, the generated sequence could be more meaningful.

\noindent\textbf{Training Details} ~ Except for Code2vec \cite{alon2019code2vec}, the other baseline models and our proposed model are all based on the encoder-decoder framework. For all of these models, the dimension of the hidden unit for both encoder and decoder, token embeddings, and summary embeddings are set to 128. The lengths of the source code and summaries are 300 and 30. Sequences that exceed the maximum lengths will be excluded from training. The vocabulary size of the code and summary are 50,000 and 26,971. To train the model, we use Adam optimizer with a base learning rate of 1e-3. The batch size is set to be 32. For Code2vec \cite{alon2019code2vec}, the AST paths are used to represent programs. We apply the same experiment setting as in Alon et al. \cite{alon2019code2vec}. Their model is originally used for method name prediction, and we use their model to encode the source code and build an LSTM network as the decoder to generate the summary. The parameters of the decoder are set the same as the other baselines. For Hybrid2Seq+Attn+DRL \cite{wan2018improving}, the parameter configuration and the training details of the actor network and critic network are set the same as in Wan et al. \cite{wan2018improving}.

\section{Result and Analysis}\label{res}
\subsection{RQ1: How does our proposed model perform when compared with baseline models?} 
To answer the first research question, we evaluate our proposed model on three tasks and compare its performance with the baseline models. 
 
 \begin{table*}
  \begin{center}
    \caption{Accuracy comparison of baseline approaches and SA-LSTM on code completion.}
    \label{tab:code_completion_acc}
    \setlength{\tabcolsep}{2mm}{
    \begin{tabular}{lcccc} 
      \hline
      ~ & \multicolumn{2}{c}{\bf{Python}} & \multicolumn{2}{c}{\bf{C}}\\
      ~ & \bf{NT} & \bf{T} & \bf{NT} & \bf{T} \\
      \hline
      Vanilla LSTM & 65.32\% & 61.53\% & 80.81\% & 73.57\% \\
      Pointer Mixture Network \cite{Li2018Code} & - & 64.36\% & - & 77.86\%  \\
      SA-LSTM & \textbf{68.23}\% & \textbf{64.54}\%  & \textbf{83.30\%} & \textbf{78.07\%} \\
      \hline
    \end{tabular}
    }
  \end{center}
\end{table*}

 \begin{table*}
  \begin{center}
    \caption{MRR comparison of baseline approaches and SA-LSTM on code completion.}
    \label{tab:code_completion_mrr}
    \setlength{\tabcolsep}{2mm}{
    \begin{tabular}{lcccc} 
      \hline
      ~ & \multicolumn{2}{c}{\bf{Python}} & \multicolumn{2}{c}{\bf{C}}\\
      ~ & \bf{NT} & \bf{T} & \bf{NT} & \bf{T} \\
      \hline
      Vanilla LSTM & 71.07\% & 67.84\% & 85.99\% & 78.52\% \\
      Pointer Mixture Network \cite{Li2018Code} & - & 69.85\% & - & 82.24\%  \\
      SA-LSTM & \textbf{75.44}\% & \textbf{71.03}\%  & \textbf{88.73\%} & \textbf{83.28\%} \\
      \hline
    \end{tabular}
    }
  \end{center}
\end{table*}

\subsubsection{Code Completion} 
The accuracies of SA-LSTM and the baseline models are presented in Table \ref{tab:code_completion_acc}, and the MRR scores are presented in Table \ref{tab:code_completion_mrr} \footnote{\textit{Pointer Mixture Network} is proposed to solve the OoV problem and there is no unknown word problem in non-terminal's prediction. Hence, there is no results in non-terminal's prediction for Pointer Mixture Network.}. We evaluate the performance of the models on predicting the next terminal and non-terminal token. In the table, \textit{NT} and \textit{T} means the results of non-terminal prediction and terminal prediction, respectively. As we have mentioned before, non-terminals contain structural information of the program, and terminals represent the semantics of programs. The results show that SA-LSTM model performs better than baseline models. The results on non-terminal tokens' prediction demonstrate our proposed model can learn the structural information of the source code, which is of great importance in code completion task. The results on terminal tokens' prediction indicate that our proposed model can also learn the program semantics well. The above results indicate that we offer a better approach for modeling programs by capturing the hierarchical structural information and can achieve better performance in code completion. 

\begin{figure*}
\setlength{\abovecaptionskip}{0cm} 
\centering\includegraphics[width=14cm]{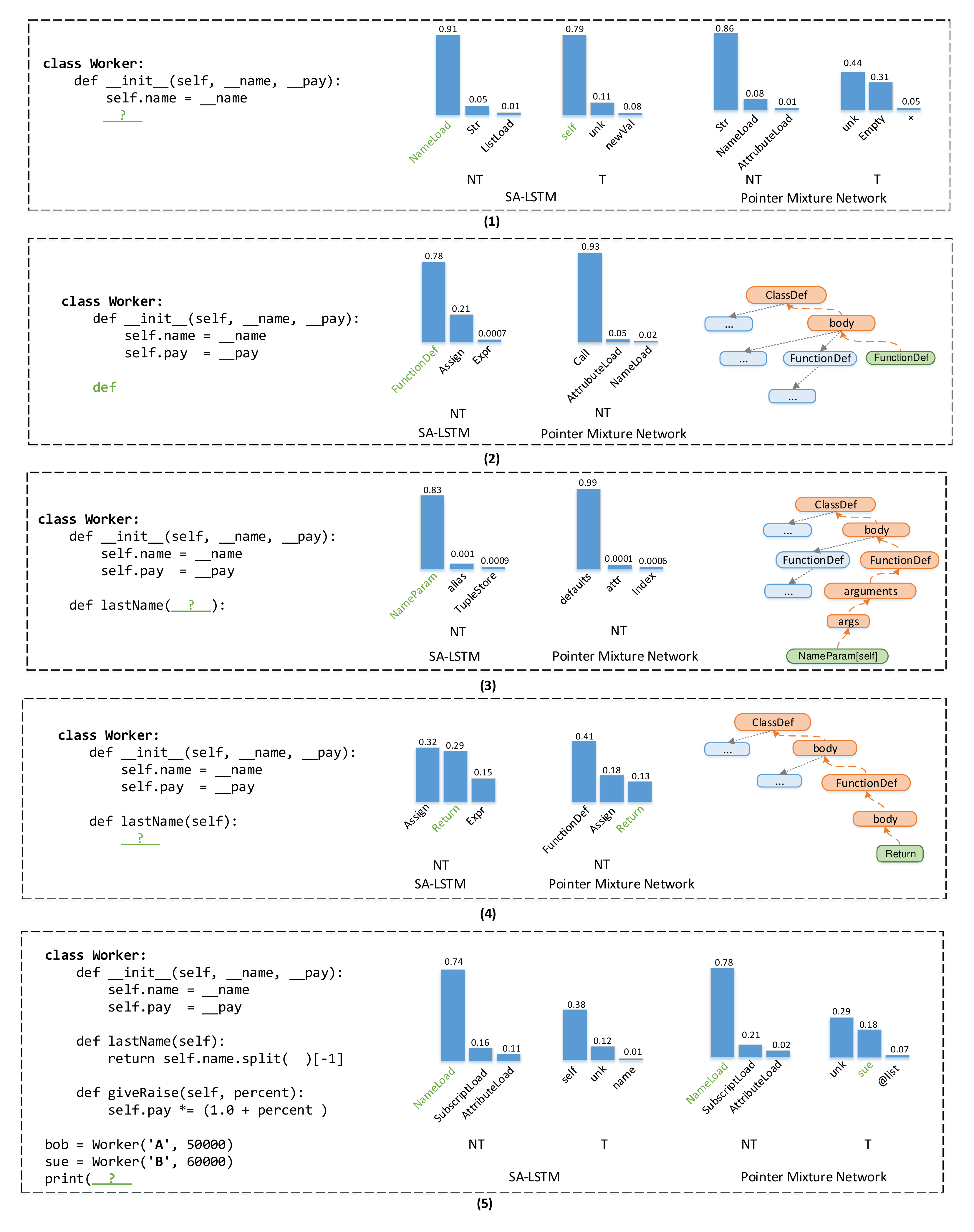}
\caption{Code completion examples} 
\label{Fig:case_study_code_completion}
\end{figure*}

\noindent\textbf{Qualitative analysis} ~ Here, we present code completion examples on a Python code snippet to analyze the performance of our proposed model qualitatively. The complete code example is shown as follows. We take several positions in the above code example to test the performance of our model. We show the top three predictions of our model and Pointer Mixture Network \cite{Li2018Code}. The results are shown in Figure \ref{Fig:case_study_code_completion}. The first three cases show the successful cases, and the last two are failures.

\lstset{frame=single}
\begin{lstlisting}[caption={A Python code example.}]
class Worker:
    def __init__(self, __name, __pay):
        self.name = __name
        self.pay  = __pay
    def lastName(self):
        return self.name.split()[-1]
    def giveRaise(self, percent):
        self.pay *= (1.0 + percent)
bob = Worker('A', 50000)
sue = Worker('B', 60000)
print(bob.lastName())
print(sue.lastName())
sue.giveRaise(.10)
print(sue.pay)
\end{lstlisting}

In the first example, the target prediction \textit{NameLoad(self)} is not a new variable and has been used in the previous context. By correctly predicting \textit{NameLoad} as its corresponding non-terminal, our model can realize the target terminal is an already used value in the previous context. Thus it can identify the value from the context and make a correct prediction. While the baseline model predicts \textit{Str} as the corresponding non-terminal, thus it might not realize the prediction is a variable accessing operation, thus predicts \textit{unk} as a result.

In the second example, the target prediction \textit{def} means a function definition, where its corresponding node is \textit{FunctionDef}. With the help of the stack, our model could discover the hierarchical structure of the contextual code, that is, by tracing the beginning and the end of the code block, the model can realize the target prediction is a new code block after a function definition block (\textit{\_\_init\_\_}), and both of them are nested in a class (\textit{Worker}). Thus, our model can assign a high probability to \textit{FucntionDef} when predicting the next non-terminal, because in a Java class, a function definition is always followed by a new function definition. From the AST perspective, the body of the class definition always contains several \textit{FunctionDef} nodes, and these nodes are neighbors. While the baseline model can only learn from the sequential context and fail to recognize the hierarchical structure of the context, thus it cannot produce the right prediction.

In the third example, the target prediction \textit{NameParam(self)} is a parameter for the function \textit{lastName}, and its corresponding non-terminal is \textit{NameParam}. By discovering the hierarchical structure of the contextual code, our model can realize the prediction is an argument of a function, thus can make the correct prediction on the non-terminal, while the baseline model fails.

In the fourth example, both of our model and the baseline model fail to produce the correct prediction \textit{Return}. In this case, the hierarchical structure of the contextual program cannot offer accurate information because there exist many possible children for a function's body. Thus, our model produces \textit{Assign}, which is also a grammatical child. The correct prediction is ranked second in our model, and is ranked third in the baseline model. In cases like this, our model might make wrong predictions.

In the last example, the target prediction \textit{sue} is a class name defined by the user. This terminal rarely occurs in the training corpus, thus is an OoV value with respect to the whole training corpus. Our model fails to predict this terminal but can correctly predict its corresponding non-terminal. The baseline model successfully identifies the OoV value from the context through the pointer network.

To sum up, compared with the baseline model, our model can take advantage of the hierarchical structure of AST, thus can achieve better results in most cases. When predicting an OoV terminal or the contextual program cannot offer determinate information for the completion, our model might fail.

\subsubsection{Program Classification}
The accuracy of program classification is shown in Table \ref{Tab:classification}. The results show that SA-LSTM model performs better than all the three baselines, where both the Tree-LSTM and TBCNN consider the structure of programs. It is worth noting that we do not use pre-trained embeddings for the AST nodes, and our model still outperforms TBCNN. In TBCNN, removing the pre-trained embeddings will lead to 2\% degradation in performance (the accuracy becomes 92.3\%), which is reported in their paper \cite{mou2016convolutional}. In Tree-LSTM, AST nodes are processed bottom-up, i.e., from children to their parent. When processing the child nodes, Tree-LSTM does not know the information about its parent node. Moreover, Tree-LSTM does not consider the order of children nodes. Thus, Tree-LSTM cannot capture the parent node's information and sequential information of siblings, so it will lose the contextual features of the programs. In TBCNN, tree-based convolution kernels are adopted to extract the structural information of programs. However, the final representation vectors for programs are computed by a pooling operation, and this process might lead to the loss of the structural information.  

\begin{table}
  \begin{center}
    \caption{Accuracy comparison of baseline approaches and SA-LSTM on program classification.}
    \label{Tab:classification}
    \setlength{\tabcolsep}{8mm}{
    \begin{tabular}{lcc}
      \hline
      \textbf{Model} & \textbf{Accuracy}\\
      \hline
       LSTM & 89.77\% \\
       Tree-LSTM & 91.46\% \\
       TBCNN & 94.0\% \\
       SA-LSTM & \bf94.78\% \\
      \hline
    \end{tabular}
    }
  \end{center}
\end{table}

In our model, we traverse an AST in pre-order, so that the parent nodes are processed before their child nodes. Thus, our model can utilize the information of the parent node when processing child nodes. After processing each parent node, our model stores its information into the stack. Then the child nodes are processed sequentially, and the sequential information of the siblings can be captured. After processing all the child nodes, our model backtracks to the parent node to restore its information stored in the stack before, and then combine it with the children's information to get a more precise understanding of the entire sub-tree. The above process (from parent to children and then back to parent) ensures that our model makes full use of the tree structure as well as the contextual features of programs, which helps the model understand programs' functionalities and produce precise representations for programs. Therefore, our proposed model can achieve better performance in this task.

\subsubsection{Code Summarization}
The BLEU-4 and METEOR scores of SA-LSTM model and the baseline model on code summarization task are presented in Table \ref{tab:summary}. It can be observed that SA-LSTM obtains outperform all the baselines in terms of BLEU-4 score, and achieves comparable performance with HAD \cite{wan2018improving} in terms of METEOR. The performance of Code2vec \cite{alon2019code2vec} is poor in the code summarization task, and even worse than vanilla seq2seq model. The reasons are as follows. They extract a collection of AST paths from the AST to represent a complete code snippet, which helps the model capture the data-flows in programs and learn how variables are used. Thus, their model mainly focuses more on the data-flow information, which is useful in method name predicting task. However, in code summary generation task, the semantic and structural information plays an important role. In Code2vec, they only use discrete AST paths to represent the code snippets, semantics and the structural information of the complete code snippets will not be fully considered, which leads to the poor performance on code summary generation. For TL-CodeSum \cite{Hu18Summarizing}, they generate summaries with the assistance of transferred API knowledge, so their model achieves better performance than the vanilla seq2seq+attention model. In HAD \cite{wan2018improving}, they also consider the structural information of the source code, and also adopt reinforcement learning during the training process. While in our model, we do not adopt reinforcement learning and can still achieve comparable performance with them.

 \begin{table}
  \begin{center}
    \caption{Results of baseline approaches and SA-LSTM on code summarization for programs.}
    \label{tab:summary}
    \setlength{\tabcolsep}{5mm}{
    \begin{tabular}{ccc} 
      \hline
      ~ & \bf{BLEU-4} & \bf{METEOR}  \\
      \hline
      Code2vec & 35.66 & 14.73 \\
      seq2seq+attention & 37.94 & 16.35 \\
      TL-CodeSum  & 41.98 & 18.81 \\
      HAD & 43.15 & \textbf{19.28} \\
      SA-LSTM+attention &  \textbf{43.94} & 19.04 \\
      \hline
    \end{tabular}
    }
  \end{center}
\end{table}

\noindent\textbf{Qualitative analysis} ~ Here, we perform qualitative analysis on the summarization generated by our model and other baseline models. We show three examples in Table \ref{tab:case_study_summary}. 

In the first example, the Java method aims to assign node weights through a for loop recursively. The summary generated by our model is even better than the ground truth because our model can understand the structure of the code better. The summaries generated by the other three baselines cannot reflect the recursive operation of the Java methods. Thus, these summaries are inaccurate. Among these baselines, the summary generated by Code2vec \cite{alon2019code2vec} is worst. The content of it is inconsistent with the Java methods, because this model concerns more about the data-flow of the code, and ignores the structural information.

In the second example, the summary generated by our model is closest to the ground truth. This Java method aims to check if all the elements in a list are real numbers. The summaries generated by the other three baselines only focus on a single element (string, number, coordinate of a vector). While our model recognizes the for loop in the code snippet, and can generate the correct comment.

In the third example, the Java method is short and easy. Both of the baseline models and our model can generate similar and correct summaries. In cases like this, the strength of our model cannot be reflected. 

These examples demonstrate that SA-LSTM can understand the structure of the programs better, thus can generate more accurate summaries for the code snippets in most cases.

\lstset{frame=none}
 \begin{table*}
  \begin{center}
    \caption{Examples of generated summaries given Java methods.}
    \setlength{\tabcolsep}{2mm}{
    \label{tab:case_study_summary}
    \begin{tabular}{l|l} 
      \toprule
      \multicolumn{2}{c}{\bf{Examples}} \\
      \midrule
      Java Method  & 
      \begin{lstlisting} 
 private void computeWeights (Node node) { 
    int wsum = _NUM; 
    for (Node child : node.children) { 
        computeWeights (child); 
        wsum += child.weight; 
    } 
    node.weight = Math.max (_NUM , wsum); 
 } \end{lstlisting} \\
 \midrule
      Groud Truth  & recursively assign node weights. \\
      seq2seq+attention & assign a value to an element. \\
      TL-CodeSum & creates an operation to assign a value to an array element. \\
      Code2vec & computes the weighted distance between the two specified vectors. \\
      HAD & assign the value to all nodes.\\
      SA-LSTM & recursively visit all nodes and set nodes weights. \\
      \bottomrule
      
        Java Method  & 
      \begin{lstlisting} 
 public static void checkFinite(final double[] val) 
 throws MathIllegalArgumentException{ 
    for(int i=_NUM;i<val.length;i++){ 
        final double x = val[i]; 
        if(Double.isInfinite(x)||Double.isNaN(x)){ 
            throw new MathIllegalArgumentException(
            LocalizedCoreFormats.NOT_FINITE_NUMBER,x); 
        } 
    } 
}\end{lstlisting} \\
 \midrule
      Groud Truth  & checks that all the elements are real numbers. \\
      seq2seq+attention & checks correctness of the given string. \\
      TL-CodeSum & checks if a number is finite. \\
      Code2vec & returns true if any coordinate of this vector is infinite. \\
      HAD & check wheather the given number is valid. \\
      SA-LSTM & determine whether all the numbers are valid. \\
\bottomrule

        Java Method  & 
      \begin{lstlisting} 
 public void removeSwipeListener(SwipeListener 
 listener){ 
    if(mListeners == null){ 
        return; 
    } 
    mListeners.remove(listener) ; 
} \end{lstlisting} \\
 \midrule
      Groud Truth  & removes a listener from the set of listeners. \\
      seq2seq+attention & removes a listener from the set of listeners. \\
      TL-CodeSum & removes a listener from the set of listeners. \\
      Code2vec & removes a listener from the list. \\
      HAD & removes a listener from the set of listeners. \\
      SA-LSTM & removes a listener from the set listening to this animation. \\
\bottomrule
    \end{tabular}
    }
  \end{center}
\end{table*}

\subsection{RQ2: How do different $\alpha$ functions impact the performance of our proposed model?}
$\alpha$ is an arbitrary trainable function which is designed for extracting the important contextual information when a complete code block has been accessed by the network. We have tried three methods based on the purpose of extracting valuable contextual information from $h_{begin}$ and $h_{t-1}$: 
\begin{itemize}
\item FC: $h_{context}=fc(concat(h_{begin},h_{t-1}))$
\item Max-pooling: $h_{context}=maxpooling(h_{begin},h_{t-1})$
\item Summarization: $h_{context}=lstm(h_{t-1},h_{begin})$
\end{itemize}

\begin{table*}
  \begin{center}
    \caption{The performance of different implementations for $\alpha$ function on code completion task.}
    \label{tab:alpha_code_completion}
    \setlength{\tabcolsep}{3mm}{
    \begin{tabular}{c|c|c|c|c} 
      \hline
      \textbf{Dataset} & \textbf{Token Type} & \textbf{FC} & \textbf{Max-pooling} & \textbf{Summarization}\\
      \hline\hline
       \multirow{2}*{Python} & NT & 65.64\% & 67.15\% & \bf{68.23\%} \\
		~ & T &	 62.54\% & 63.92\% & \bf{64.54\%} \\
      \hline\hline
        \multirow{2}*{C} & NT & 77.18\% & 77.65\% & \bf{83.30\%} \\
		~ & T &	 65.19\% & 65.95\% & \bf{78.07\%} \\
      \hline
    \end{tabular}
    }
  \end{center}
\end{table*}

\begin{table}
  \begin{center}
    \caption{The performance of different implementations for $\alpha$ function on code classification task.}
    \label{Tab:alpha_classification}
    \setlength{\tabcolsep}{8mm}{
    \begin{tabular}{lcc}
      \hline
      \textbf{$\alpha$} & \textbf{Accuracy}\\
      \hline
       FC & 90.34\% \\
       Max-pooling & 93.97\% \\
       Summarization & \bf94.78\% \\
      \hline
    \end{tabular}
    }
  \end{center}
\end{table}

 \begin{table}
  \begin{center}
    \caption{The performance of different implementations for $\alpha$ function on code summarization task.}
    \setlength{\tabcolsep}{3mm}{
    \label{tab:alpha_summary}
    \begin{tabular}{lcc} 
      \hline
      $\alpha$ & \bf{BLEU-4} & \bf{METEOR}  \\
      \hline
      FC & 39.88 & 14.95 \\
      Max-pooling & 43.81 & 18.97 \\
      Summarization  & \textbf{43.94} & \textbf{19.04}  \\
      \hline
    \end{tabular}
    }
  \end{center}
\end{table}

We evaluated these methods on code completion, program classification, and code summarization task. The results is presented in Table \ref{tab:alpha_code_completion}, Table \ref{Tab:alpha_classification}, and Table \ref{tab:alpha_summary}. The evaluation results demonstrate the last method achieves better results than the other two for all the three tasks. For program classification and code summarization task, Max-pooling achieves comparable results with Summarization method. The reasons for the good performance of Summarization method are as follows. In this method, when the model has read all the tokens in a code block, by setting the current hidden state to $h_{begin}$, the model would forget the details about the code block just like they have not seen them. Then taking $h_{t-1}$ as input, which contains the key information of the forgotten part, our model can store this information into the new hidden state and then continue the learning process. This process enables our model to extract the important information of the contextual code block, thus can achieve better performance. The other two methods use a fully-connected layer or a max-pooling layer to combine $h_{begin}$ and $h_{t-1}$, which are insufficient to extract the important features from the contextual code.  

\begin{figure}
\centering\includegraphics[width=7cm]{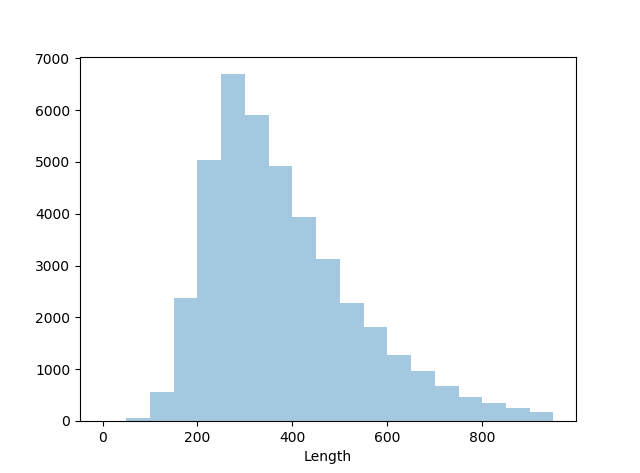} 
\caption{The distribution of the programs' length on the program classification dataset.} 
\label{Fig:length}
\end{figure}

\subsection{RQ4: Is our model effective for capturing long-term dependency in programs?} \label{long_dependency}
In our model, we augment the LSTM network with a stack to store and restore the contextual information depending on the programs' hierarchical structure, which enables the model to capture the structural information of the programs. We conjecture that understanding the structure of programs helps our model capture longer-term dependency than the baseline models. To verify this conjecture, we conduct experiments on program classification task under different program length settings. The reasons for using the program classification task for verification are as follows. In code completion task, each token in the programs needs to be predicted. When the program's length varies, the tokens which are used in training and test process also varies. Thus, it is hard to evaluate the model's ability of capturing the long-term dependency from the results. For the code summarization task, the length of the logical representations is small, so it is not sufficient to evaluate the ability to capture the long dependency. Hence, we use the program classification task to conduct experiments. The distribution of the programs' length on the program classification dataset is shown in Figure \ref{Fig:length}.

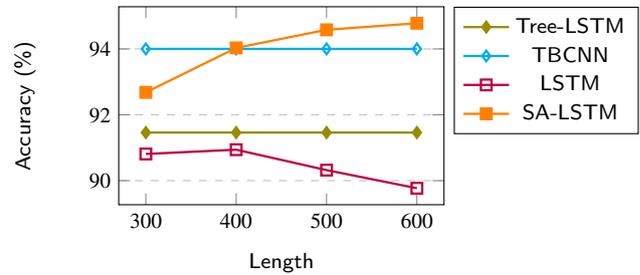
\begin{figure}
\begin{flushleft}
\begin{tikzpicture}
\footnotesize
\begin{axis}[width=0.7\columnwidth,
            height=0.5\columnwidth,
            xlabel=Length,
            ylabel=Accuracy (\%),
            xtick={300,400,500,600},
            legend pos= outer north east,
            ymajorgrids=true,
            grid style=dashed]
\addplot[line width=0.3mm, mark=diamond*, color=olive] coordinates {(300,91.46)(400,91.46)(500,91.46)(600,91.46)};
\addlegendentry{Tree-LSTM}
\addplot[line width=0.3mm, mark=diamond, color=cyan] coordinates 
{(300,94.0)(400,94.0)(500,94.0)(600,94.0)};
\addlegendentry{TBCNN}
\addplot[line width=0.3mm, mark=square,color=purple] coordinates {(300,90.81)(400,90.94)(500,90.32)(600,89.77)};
\addlegendentry{LSTM}
\addplot[line width=0.3mm, mark=square*,color=orange] coordinates {(300,92.68)(400,94.03)(500,94.58)(600,94.78)};
\addlegendentry{SA-LSTM}
\end{axis}
\end{tikzpicture}
\end{flushleft}
\caption{Accuracy of baseline approaches and SA-LSTM on program classification under different length settings.}
\label{Fig:len_accuracy}
\end{figure}

As shown from Figure \ref{Fig:length}, the lengths are mainly distributed in the range from 300 to 600. We conduct program classification experiments varying the program length on LSTM and our proposed model. The other two baselines (Tree-LSTM and TBCNN) take the full program AST as inputs, and the programs are not truncated. The results are shown in Figure \ref{Fig:len_accuracy}. 

As seen from the results, for the standard LSTM, when increasing the length from 300 to 400, the accuracy increases a little. As the length continues to increase, the accuracy begins to decrease. Thus, we can get the conclusion that the maximum valid memory length of the LSTM is 400 in this experiment, which is due to the vanishing gradient problem of the LSTM network. And this problem prevents the model from capturing long-term dependency of the input data. While in SA-LSTM, as the input program length increases, the accuracy increases, and the growth trend slows down from 500 to 600. It is worth noting that when the input program length is 300, the performance of TBCNN \cite{mou2016convolutional} is better than our model. Actually, this comparison is unfair. In TBCNN, they take the full program AST as the input. The results of TBCNN are produced by conducting experiments on the complete ASTs, and the programs are not truncated. While in our experiments, much information will be lost after truncating the programs. When the program length increases, the information loss is reduced, our model can achieve better performance than TBCNN. In this dataset, the programs are from an open judge platform, where most of the programs are simple and short, and the highest distribution length (of the AST nodes) is around 300. In actual use, programs are always more complex and longer than this dataset. Thus, our model can play its advantages by capture the structural information and the long-term dependency in programs.

The above results demonstrate that by capturing the hierarchical structural information, our proposed model can capture longer-term dependency of the input programs.  

\section{Discussion}\label{discussion}
In this section, we first discuss some limitations of our model and potential future research directions. Then we present some threats to validity.

\subsection{Limitations of Our Model}
\noindent\textbf{Closed vocabulary} ~ One of the major limitations of our model is the closed vocabulary. In our model, we limit vocabulary to the most common K (e.g., 5,000) tokens in a pre-processing step. Words outside this vocabulary are treated as OoV tokens, and these tokens cannot be predicted by a language model. If many OoV tokens exist in a program, it will bring lots of difficulties for understanding the program semantic. Overall, our model can still outperform the Pointer Mixture Network \cite{Li2018Code}, which adopts a pointer network to ease the OoV issue. In the future, we plan to extend our model to an open-vocabulary language model by representing the tokens as sub-word units. In this way, an OoV token that has never occurred in the training data can still be represented as su-bword units will have occurred in training, and the vocabulary size will become much smaller.

\noindent\textbf{Computation efficiency} ~ We adopt LSTM, a variant of the recurrent neural network, as the base language model. In recurrent models, each hidden state $h_t$ is computed depending on the previous hidden state $h_{t-1}$. Lots of recurrent computations are performed during the hidden state updating process, which leads to much computational complexity in the training process when compared to the traditional non-neural network based language models, which is the cost of the improvements. In the future, we plan to change our model architecture to self-attention \cite{vaswani2017attention} based neural networks, which can improve the computational efficiency by parallelizable computations across sequence positions.

\subsection{Threats to Validity}
\noindent\textbf{Threats to external validity} relate to the quality of the datasets we used and the generalizability of our results. For code completion and program classification, the programs of the datasets are collected from GitHub repositories or OJ system. For code summarization, we perform experiments on java dataset, where the programs are also collected from GitHub repositories. Our corpus is representative in practice, and the commonality we have seen across these different datasets gives us confidence that our results hold generally. 

\noindent\textbf{Threats to internal validity} include the influence of the hyper-parameters used in our model. The performance of our model would be affected by different hyper-parameter settings, which are tuned empirically in our experiments. Thus, there is little threat to the hyper-parameter choosing, and there might be room for further improvement. However, current settings have achieved a considerable performance increase. Another threat to internal validity relates to the representation of the programs. In this work, the hierarchical structure of the programs is inferred through a supervised syntactic parser. Programs are first parsed into ASTs, and then we extract the hierarchical features from ASTs. If syntax errors exist in programs, or there are not supervised syntactic parsers, the programs can not be parsed into ASTs. However, ASTs have been widely used in program processing \cite{raychev2016probabilistic,bielik2016phog,Li2018Code,alon2019code2vec}, and are effective for representing the programs' structure. For most of the widely used programming languages, supervised syntactic parsers are provided. In the future, learning the tree structure from the programs in an unsupervised manner would be a potential direction.

\noindent\textbf{Threats to construct validity} relate to the suitability of our evaluation measure. For the code completion task, we use accuracy and MRR as the metrics which evaluate the proportion of correctly predicting the next token. These metrics are classical evaluation measures for code completion, and have been used in many previous code completion models \cite{hellendoorn2017deep,hindle2012naturalness,raychev2016probabilistic,Li2018Code}. For the program classification task, we use accuracy as the metric to evaluate the proportion of correctly predicting the program's categories, which has been used in previous program classification work \cite{mou2016convolutional}.
For the code summarization task, we use BLEU and METEOR as metrics to evaluate the quality of the summary by measuring the n-gram overlap between the summaries produced by models and the reference summaries. These two metrics are widely used in code summarization work \cite{iyer2016summarizing,Hu18Summarizing,wan2018improving}.

\section{Conclusion}\label{conclusion}
The hierarchical structure of programs is critical for many program analysis tasks. To this end, we propose a Stack-Augmented LSTM model that can learn the hierarchical structure of programming languages. We use AST to represent and preserve the structure of programs. To enable the model to learn this information, we strengthen the standard LSTM network by adding a stack as a memory unit to trace the beginning and end of code blocks, store and restore contextual information depending on the hierarchical structure of the programs. We evaluate the proposed model on three program analysis tasks. Evaluation results suggest that adding a stack as a memory component enables SA-LSTM to extract the hierarchical features of programming languages and capture long-range dependency in programs, which eventually helps the model understand programs and provide a more accurate representation for programs.

\section*{Acknowledgements} This research is supported by the National Key R\&D Program under Grant No. 2018YFB1003904, and the National Natural Science Foundation of China under Grant No. 61832009, and is partially supported by the National Natural Science Foundation of China under Grant No. 61620106007 and No. 61751210. Zhi Jin is the corresponding author.

\bibliographystyle{cas-model2-names}

\bibliography{ref}

\end{document}